\documentclass[3p,times]{elsarticle}

\usepackage{ecrc}

\volume{00}

\firstpage{1}

\journalname{Current Opinion in Electrochemistry}

\runauth{}

\jid{}

\jnltitlelogo{}

\CopyrightLine{2021}{Published by Elsevier Ltd.}

\usepackage{amssymb}

\usepackage[figuresright]{rotating}
\usepackage{subfig}

\begin{document}

\begin{frontmatter}



\dochead{}

\title{Electric double layer theory for room temperature ionic liquids on charged electrodes: milestones and prospects}

\author[label1,label2]{Yury A. Budkov}
\address[label1]{School of Applied Mathematics, HSE University, Tallinskaya st. 34, 123458 Moscow, Russia}
\address[label2]{G.A. Krestov Institute of Solution Chemistry of the Russian Academy of Sciences, 153045, Akademicheskaya st. 1, Ivanovo, Russia}
\ead{ybudkov@hse.ru}
\author[label3]{Andrei L. Kolesnikov}
\address[label3]{Institut f\"ur Nichtklassische Chemie e.V., Permoserstr. 15, 04318 Leipzig, Germany}
\begin{abstract}
In this review, we shortly summarize the basic theoretical milestones achieved in the mean-field theory of room temperature ionic liquids (RTILs) on charged electrodes since the publication of Kornyshev's seminal paper in 2007. We pay special attention to the behavior of the differential capacitance profile and the microscopic parameters of ions that can have substantial influence on it. Among them are parameters of short-range specific interactions, ionic diameters, static polarizabilities, and permanent dipole moments. We also discuss the recent "nonlocal" mean-field theories that can describe the overscreening behavior of the local ionic concentrations, as well as the crossover from overscreening to crowding.
\end{abstract}


\end{frontmatter}

\begin{figure}[h]
\centering
\includegraphics[width=0.75\linewidth]{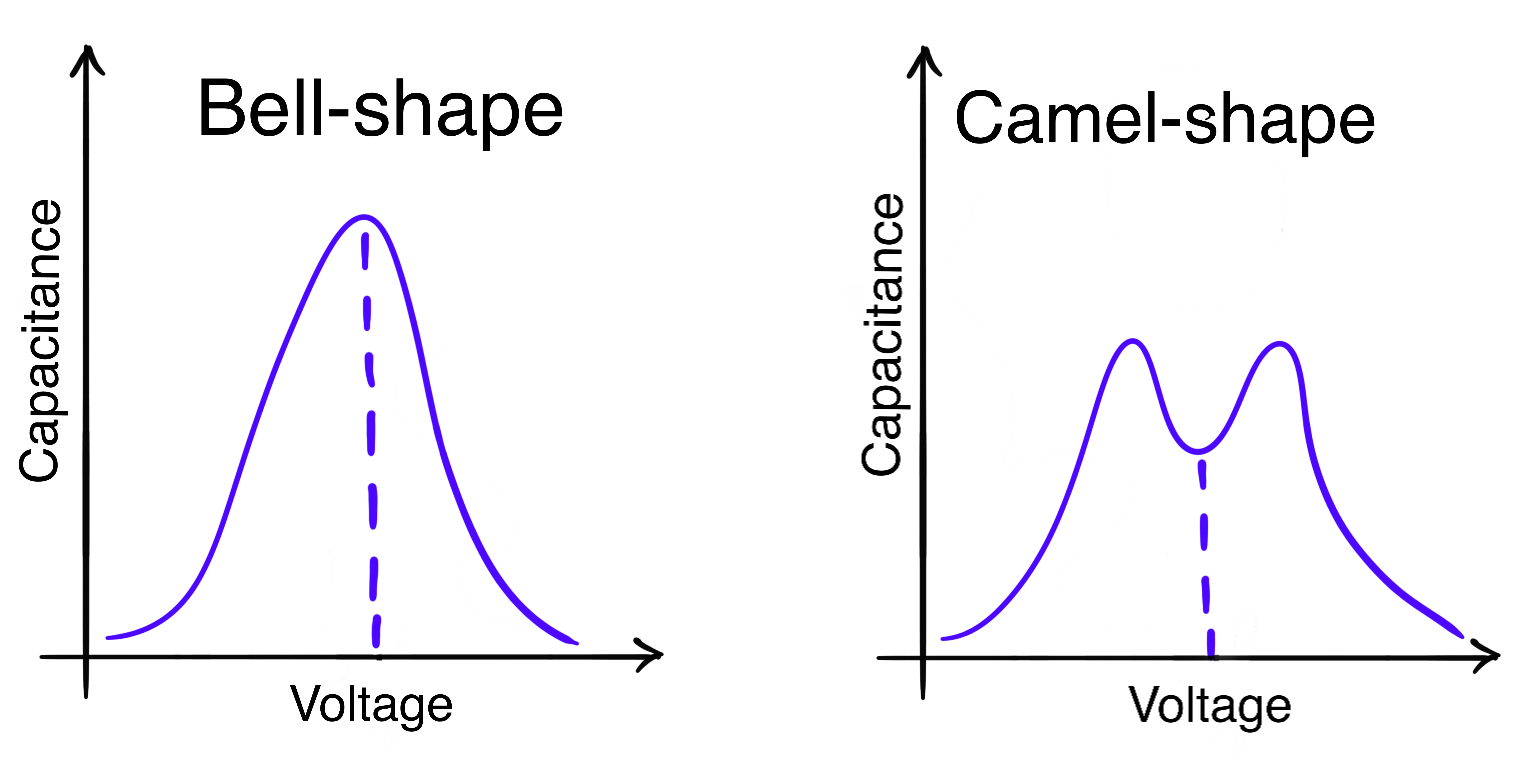}
\caption{Schematic representation of the typical shapes of differential capacitance profiles.}
\label{fig1} 
\end{figure}

{\bf Introduction}.
The Gouy-Chapman (GC) theory of the diffusive electrical double layer (EDL) in electrolyte solutions at the metal-liquid electrolyte interface, based on the application of the classical Poisson-Boltzmann (PB) equation, was successfully applied to dilute electrolyte solutions, but failed to describe the EDL of room temperature ionic liquids (RTILs are molten salts that are liquid around room temperature~\cite{walsh2018oxygen}) on charged electrodes. The U-shape profile of the EDL differential capacitance (DC), predicted by the GC theory, is in total disagreement with its experimentally observed behavior (camel-shaped or bell-shaped profiles). This problem had remained unsolved until 2007 when Kornyshev proposed a generalization of the GC theory~\cite{kornyshev2007double}, based on the ideal symmetric lattice gas model~\cite{hill1986introduction} (each lattice cell cannot store more than one ion), which takes into account the excluded volume of the ions, in contrast to the ideal ionic gas model, utilized in the GC theory. Solving the self-consistent field equation
\begin{equation}
\varepsilon\varepsilon_0\psi^{\prime\prime}(z)=\frac{2c q\sinh{\beta q\psi(z)}}{1+2cv\left(\cosh\beta q\psi(z)-1\right)}
\end{equation}
for the electrostatic potential $\psi(z)$ (the so-called Poisson-Fermi (PF) equation~\cite{borukhov1997steric,bazant2009towards}) with the boundary conditions $\psi(0)=\psi_0$, $\psi^{\prime}(\infty)=0$, Kornyshev obtained a simple analytical expression for the DC as a function of voltage $\psi_{0}$
\begin{equation}
\label{Korn_cap}
C=C_{K}(\psi_0)=\frac{\left({2\beta\varepsilon\varepsilon_0c^2vq^2}\right)^{1/2}|\sinh{\beta q\psi_0}|}{\left(1+2\gamma\sinh^2\left(\frac{\beta q\psi_0}{2}\right)\right)\sqrt{\ln\left(1+2\gamma\sinh^2\left(\frac{\beta q\psi_0}{2}\right)\right)}}.
\end{equation}
where $\beta=(k_{B}T)^{-1}$, $k_{B}$ is the Boltzmann constant, $T$ is the temperature, $v$ is the lattice gas cell volume, $q$ is the charge of the ions, $\gamma=2cv$ is the volume fraction of the ions, $c$ is the bulk ion concentration, $\varepsilon$ is the RTIL dielectric constant, $\varepsilon_0$ is the vacuum permittivity. Note that eq. (\ref{Korn_cap}) predicts the experimentally observable behavior of the DC (see Fig.\ref{fig1}). As follows from the obtained expression, at a sufficiently low ionic volume fraction in the bulk one observes a camel-shaped DC profile, whereas at sufficiently high volume fractions ($\gamma >1/3$) -- a bell-shaped one. It is important to note that in the case of the camel-shaped profile at small electrode potential values one obtains the GC behavior of the DC. However, at a certain threshold value of the potential (saturation potential, in which the near-electrode layer is completely occupied by ions) DC exceeds the maximum and then decreases at rather large voltages by the power law $C\sim |\psi_0|^{-0.5}$ (crowding regime). In the case of the bell-shaped profile, the DC has the maximum value at zero voltage and decreases symmetrically, exhibiting the same asymptotic behavior. Then, the predicted DC profiles as a function of the ion volume fraction in the bulk phase and its limiting behavior were confirmed experimentally (see, for instance, \cite{alam2007measurements}), within the molecular dynamics (MD) simulation~\cite{fedorov2008ionic,fedorov2014ionic} and by complex numerical calculations within the classical density functional theory (cDFT)~\cite{jiang2011density,wu2011classical,forsman2011classical,frydel2012close}. However, despite the triumph of such simple analytical theory, it contains a number of fundamental limitations, preventing it from being utilized for the description of real RTIL on real electrodes. Firstly, despite the correct prediction of the DC profile, the predicted numerical values strongly overestimate the experimental observations. The fact that it is impossible to reach agreement between the theoretical and experimental DC values is due to the lack of free parameters in the theory that could make it possible to distinguish the ions by their size and specific interactions with each other and with the electrode. Secondly, Kornyshev's theory takes into account the electric polarization of the RTIL by introducing the effective dielectric constant, which makes it impossible to account for the specific dielectric properties of the ions (polarization and dipole moment) or to study their influence on the DC behavior. Finally, the results of the MD simulations~\cite{fedorov2008ionic} and cDFT calculations~\cite{wu2011classical} at rather small voltages demonstrate damping oscillations of the local densities and electrostatic potential near the electrode surface (so-called overscreening regime), whereas Kornyshev's theory predicts their monotonic behavior. The present review discusses the results achieved in the statistical theory of the EDL at the metal-RTIL interface since 2007. The improvements made to the theory are in one way or another related to overcoming the limitations of Kornyshev's theory. We would like to note that in present review we focus only on achievements in applications of self-consistent field theory to RTILs on charged electrodes, leaving aside the applications of cDFT and MD. We would like to mention important cDFTs including a non-central placement of charge~\cite{lu2018classical,lu2018ionic}, within a spherical ion excluded volume, but also non-spherical descriptions of anions and cations~\cite{turesson2014classical}. Comprehensive discussion of applications of cDFT and MD to RTILs undoubtedly deserves a separate review. 

\begin{figure}[h]
\centering
\includegraphics[width=0.75\linewidth]{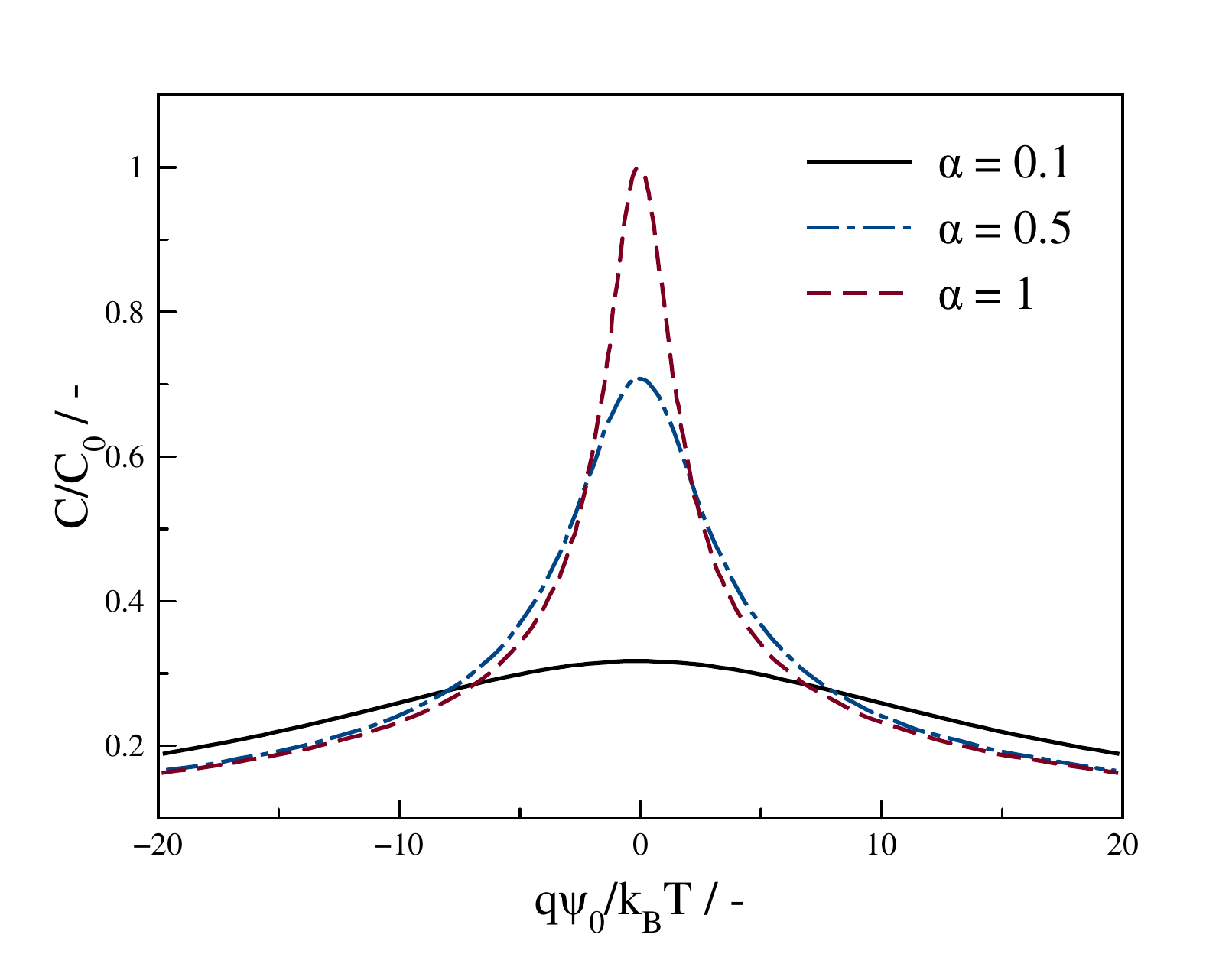}
\caption{Differential capacitance profiles calculated using eq. (\ref{Korn_cap2}), $C_0$ is the Debye capacitance. The data is taken from Ref. \cite{goodwin2017mean}.}
\label{fig2} 
\end{figure}

{\bf Short-range specific interactions and ion size asymmetry}. As it was noted in the introduction, although Kornyshev's theory qualitatively predicts the DC behavior, it overestimates the capacitance values and leads to rather sharp maxima on its voltage dependencies. Moreover, the theory does not include any adjustable parameters, whose variation could allow one to fit the experimental DC profiles. Despite the fact that the effect of size asymmetry was discussed already in Kornyshev's seminal paper, a successful attempt to modify Kornyshev's theory was firstly made in work~\cite{goodwin2017mean}, where the authors took into account additional short-range specific interactions between the ions by introducing additional quadratic terms on ionic concentrations, $c_{\pm}$, in the Bragg-Williams approximation (see, for instance, \cite{hill1986introduction}) to the Helmholtz free energy density of the lattice gas, i.e.
\begin{equation}
f=f_{0}+\frac{z_c v}{2}\left(A_{+}c_{+}^2+A_{-}c_{-}^2+2A_{+-}c_{+}c_{-}\right),
\end{equation}
where $f_{0}(T,c_{+},c_{-})=v^{-1}k_{B}T\left(\phi_{+}\ln\phi_{+}+\phi_{-}\ln\phi_{-}+(1-\phi_{+}-\phi_{-})\ln\left(1-\phi_{+}-\phi_{-}\right)\right)$ is the free energy density of the symmetric lattice gas, $\phi_{\pm}=c_{\pm}v$ are the ion volume fractions, $z_c$ is the lattice coordination number, $A_{+}$,$A_{-}$, $A_{\pm}$ are the energetic parameters of specific cation-cation, anion-anion and cation-anion interactions, respectively, which are positive or negative values (it depends on chemical specifics of ionic species). Note that the specific interactions could be related to hydrogen bonding, donor-acceptor interaction, pi-stacking, steric interactions, related to ion shape anisotropy, {\sl etc}. Basing on the linear response theory, which shows the renormalization of the Debye radius by the presence of ion specific interactions (so-called underscreening~\cite{goodwin2017underscreening,may2019differential}), the authors proposed an interpolation expression for the DC as a voltage function, comprising the dependence on the specific interaction parameters
\begin{equation}
\label{Korn_cap2}
C=\sqrt{\alpha}C_{K}(\alpha \psi_0),
\end{equation}
where $\alpha=(1+\gamma(a-b)/2)^{-1}$, $a=a_{+}=a_{-}=z_c A_{\pm}/k_{B}T$, $b=z_c A_{+-}/k_{B}T$. The authors demonstrated that the variation of $a$ and $b$ parameters in eq. (\ref{Korn_cap2}) allows one to obtain reasonable DC values and make the DC peaks on voltage dependencies smoother (see Fig.\ref{fig2}). A similar theory taking into account short-range specific interactions within quasichemical approximations was formulated in paper~\cite{downing2018role}.

In work \cite{maggs2016general} the authors present a generalization of Kornyshev's theory for the case of asymmetric effective sizes of the ions. Utilizing the local Legendre transformation, the authors introduce a grand thermodynamic potential (GTP) (per electrode unit area) of the RTIL on a flat infinite electrode as the following functional of electrostatic potential
\begin{equation}
\label{func}
\Omega[\psi]=-\int\limits_{0}^{\infty}dz\left(\frac{1}{2}\varepsilon\varepsilon_0{\psi^{\prime}}^2+P(T,\bar{\mu}_{+},\bar{\mu}_{-})\right),
\end{equation}
where $\bar{\mu}_{\pm}=\mu_{\pm}-q_{\pm}\psi$ are the intrinsic chemical potentials of the ions, $q_{\pm}=\pm q$ are the ionic charges, $\mu_{\pm}$ are the chemical potentials of the ions in the bulk, $P=P(T,\bar{\mu}_{+},\bar{\mu}_{-})$ is the local ionic pressure as a function of chemical potentials and temperature. Variation of the functional $\Omega[\psi]$ with respect to the electrostatic potential, $\psi(z)$, leads to the self-consistent field equation, which generalizes the classical PB equation
\begin{equation}
\label{scf_eq}
-\varepsilon\varepsilon_0\psi^{\prime\prime}(z)=q_{+}\bar{c}_{+}(z)+q_{-}\bar{c}_{-}(z),
\end{equation}
where $\bar{c}_{\pm}(z)=\partial{P}/\partial{\bar{\mu}_{\pm}}$ are the equilibrium local concentrations of the ions, which depend on $z$ coordinate ($z$-axis is perpendicular to the electrode with the origin on it) via the intrinsic chemical potentials $\bar{\mu}_{\pm}$. Note that for the ideal gas model, for which $P(T,\mu_{+},\mu_{-})=k_{B}T\left(\lambda_{+}^{-3}e^{\beta\mu_{+}}+\lambda_{-}^{-3}e^{\beta\mu_{-}}\right)$ ($\lambda_{\pm}$ are the thermal de Broglie wavelengths of the ions), we arrive at the PB equation. For the case of the ideal symmetric lattice gas model, for which $P(T,\mu_{+},\mu_{-})=k_{B}Tv^{-1}\ln(1+e^{\beta\mu_{+}}+e^{\beta\mu_{-}})$, we obtain the aforementioned PF equation using the expression for the bulk chemical potentials $\mu_{\pm}=k_{B}T\ln\left(cv/(1-2cv)\right)$. With the help of the developed formalism Maggs and Podgornik took into account the difference in the effective sizes of the ions within the framework of the asymmetric lattice gas model and the model of the hard spheres mixture in the Carnahan-Starling approximation. Considering the first integral of self-consistent field equation (\ref{scf_eq}) and the expression for the surface charge density of the electrode $\sigma=-\varepsilon\varepsilon_0\psi^{\prime}(0)$, they obtained a general mean-field expression for the DC as a function of voltage  
\begin{equation}
\label{cap2}
C=\frac{d\sigma}{d\psi_0}=\frac{\varepsilon\varepsilon_0|q_{-}{c}_{-s}+q_{+}{c}_{+s}|}{\sqrt{2\varepsilon\varepsilon_0\left(P_{s}-P_{0}\right)}},
\end{equation}
where $c_{\pm s}(\psi_0)=\bar{c}_{\pm}(0)$ are local ionic concentrations on the electrode, $P_{s}=P(T,\mu_{+}-q_{+}\psi_0,\mu_{-}-q_{-}\psi_{0})$ is the local particle pressure on the electrode, $P_0=P(T,\mu_{+},\mu_{-})$ is the pressure in the bulk. Notice that for the case of the symmetric lattice gas model eq. (\ref{cap2}) gives Kornyshev's DC expression (\ref{Korn_cap}). Applying expression (\ref{cap2}), it was shown that the difference in the ionic sizes for both models (based on asymmetric lattice gas and Carnahan-Starling equations of state) causes asymmetry of the camel-shaped DC profile maxima and the maximum shift for the case of the bell-shaped profile. We would also like to mention ref.\cite{han2014mean}, where Han et al. generalized the Kornyshev's model for the case of asymmetric lattice gas model and obtained more general analytical expression for the DC. However, in contrast to the approach of Maggs and Podgornik, an approach proposed by Han et al. cannot be used for the case of an off-lattice reference equations of state, such as Carnahan-Starling and Percus-Yevick equations. It is important to note also paper~\cite{yin2018mean}, where the authors generalized the theory formulated by Han et al. taking into account the short-range specific interactions of ions with Bragg-Williams approximation. In paper~\cite{zhang2018treatment} Zhang and Huang compared six lattice gas models used to account for the ion size asymmetry in EDL theory.

It is important to mention paper~\cite{budkov2018theory}, where the authors modified the theory proposed in work \cite{goodwin2017mean}, using the aforesaid formalism based on the Legendre transformation, for the case when the RTIL bulk contains a small amount of water. In this paper, the authors gave a comprehensive theoretical description of water electrosorption from the RTIL bulk to the charged electrode~\cite{bi2018minimizing,feng2014water}. Notice that that the mean-field theories of the electrolyte solutions on charged electrode with small admixture of a polarizable cosolvent were previously presented in~\cite{budkov2015modified,budkov2016theory}. We would also like to mention the paper \cite{podgornik2018general}, where with the approach of Maggs and Podgornik the ion-surface interaction effect was properly included to the electric double layer theory.

\begin{figure}[h]
\centering
\includegraphics[width=0.75\linewidth]{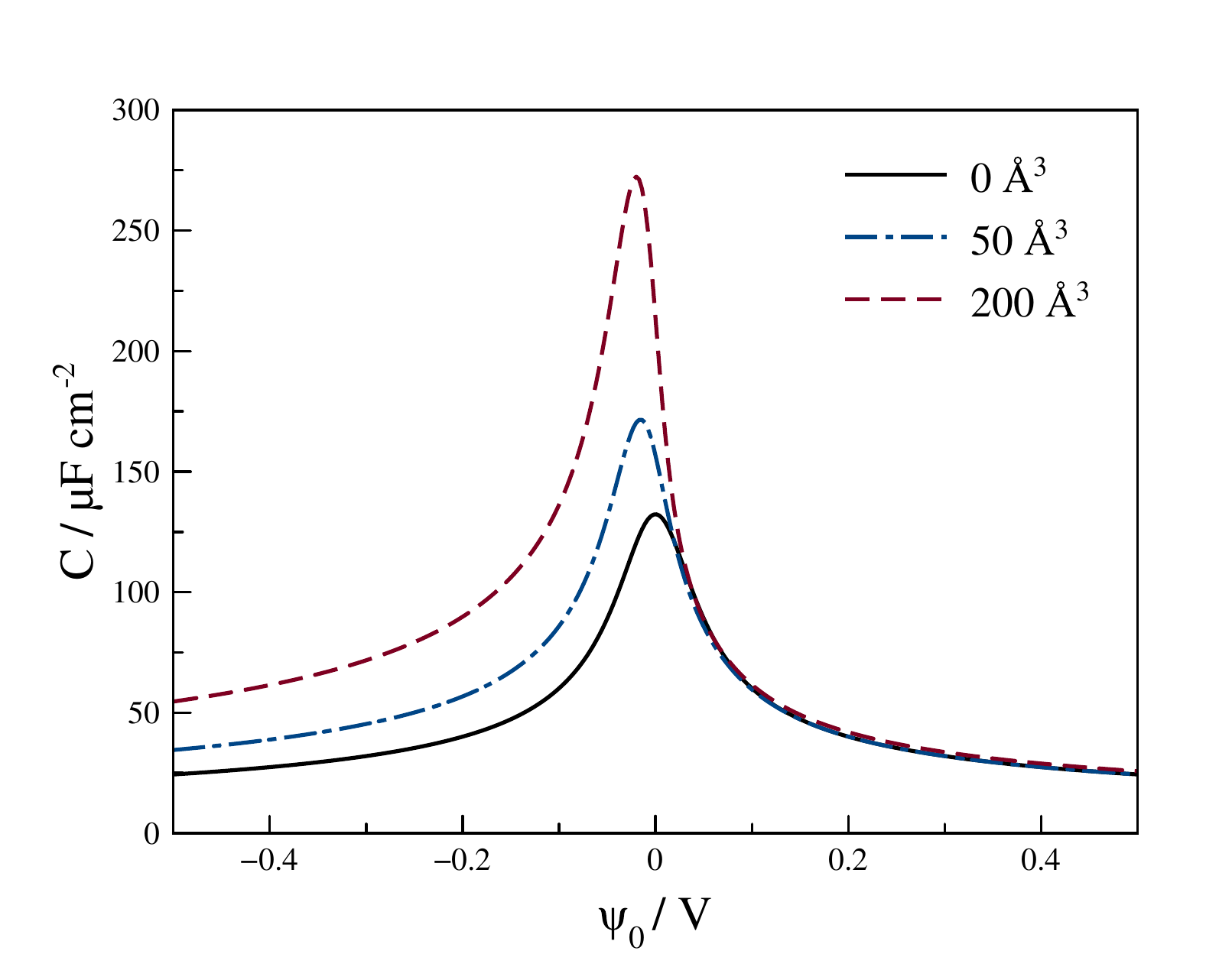}
\caption{Differential capacitance profiles of a room temperature ionic liquid with polarizable cations. The differential capacitance maximum grows and shifts to the region of positive voltages with an increase in the cation static polarizability. The same behavior is observed following an increase in the cation permanent dipole moment \cite{budkov2021theory}. The data is taken from Ref. \cite{budkov2021theory}.}
\label{fig3} 
\end{figure} 

{\bf Dipole moment and polarizability of ions.} Most of RTILs consist of inorganic weakly polarizable anions and organic molecular cations, which possess not only an electric charge but also a sufficiently high dipole moment or static polarizability~\cite{schroder2012comparing,mcdaniel2018influence,izgorodina2009components}. Polarizable organic cations, located near the charged electrode are exposed not only to the Coulomb repulsion or attraction but also to the dielectrophoretic attraction~\cite{jones1979dielectrophoretic}, acting on the permanent or induced dipole moment in the inhomogeneous electric field. Thus, presence of a sufficiently large dipole moment or static polarizability of the cations must affect the number of ions in the EDL at different voltages, and, thereby, the DC magnitude. A comprehensive study has been recently made of the polarizability and permanent dipole moment of the RTIL ions within the mean-field theory~\cite{budkov2021theory}. It is assumed that generally, ions possess charges $q_{\pm}$, permanent dipole moments $p_{\pm}$, and static polarizabilities $\alpha_{\pm}$. Based on the aforementioned Legendre transformation formalism, the authors wrote the GTP as the same functional (\ref{func}) with the reference dielectric constant, $\varepsilon$, determined by the unaccounted effects, such as formation of the ionic clusters~\cite{feng2019free,zhang2020enforced,avni2020charge} (see also below); as above, the intrinsic chemical potentials are $\bar{\mu}_{\pm}=\mu_{\pm}-q_{\pm}\psi+\Psi_{\pm}$ with the auxiliary functions $\Psi_{\pm}(z)=\alpha_{\pm}{\psi^{\prime}}^2(z)/2+k_{B}T\ln\left({\sinh{\beta p_{\pm}\psi^{\prime}(z)}}/\beta p_{\pm}\psi^{\prime}(z)\right)$. Variation of the GTP led to the following self-consistent field equation
\begin{equation}
\label{scfe2}
-\frac{d}{dz}\left(\epsilon(z)\psi^{\prime}(z)\right)=q_{+}\bar{c}_{+}(z)+q_{-}\bar{c}_{-}(z),
\end{equation}
where 
\begin{equation}
\epsilon(z)=\varepsilon\varepsilon_0+\sum\limits_{i=\pm}^{}\left(\alpha_{i}+\frac{p_{i}^2}{k_{B}T}\frac{L(\beta p_{i}\mathcal{E}(z))}{\beta p_{i}\mathcal{E}(z)}\right)\bar{c}_{i}(z)
\end{equation}
is RTIL effective local dielectric constant, taking into account the effects of the electron and orientational polarizability of the ions; $L(x)=\coth{x}-1/x$ is the Langevin function; as in the case of nonpolarizable ions, the equilibrium local concentrations $\bar{c}_{\pm}$ are dependent on the coordinates via the intrinsic chemical potentials $\bar{\mu}_{\pm}(z)$. For the bulk RTIL, where $c_{\pm}=c$, one obtains the following expression for the RTIL dielectric constant
\begin{equation}
\label{bulk_perm}
\epsilon=\varepsilon\varepsilon_0+\sum\limits_{i=\pm}^{}\left(\alpha_{i}+\frac{p_{i}^2}{3k_{B}T}\right)c.
\end{equation}
Note that the dipole moments, static polarizabilities, and reference dielectric constant can be considered as fitting parameters for the approximation of the experimental values of the RTIL dielectric permittivities in a way similar to that of the electrolyte solution theory with an explicit account of the solvent~\cite{abrashkin2007dipolar,iglivc2010excluded,gongadze2011langevin}. It is significant that for the case of the ideal ionic gas reference model, self-consistent field equation (\ref{scfe2}) transforms into the known polarizable PB equation~\cite{frydel2011polarizable}. We also mention that similar formalism~\cite{budkov2020two} was also utilized to describe a two-component electrolyte solution on a charged electrode, where one of the cations possesses a sufficiently large dipole moment. Using the first integral of the self-consistent field equation, which is reduced to the condition of the RTIL mechanical equilibrium, and the expression for the surface charge density $\sigma=\epsilon(0)\mathcal{E}(0)$, the authors obtained the following analytical expression for the DC 
\begin{equation}
\label{cap3}
C=\left|\frac{\rho_s}{\mathcal{E}_{s}}\right|,
\end{equation}
where $\mathcal{E}_s=\mathcal{E}_s(\psi_0)$ is the local electric field on the electrode, which can be found from the condition of the mechanical equilibrium on the electrode (eq. (13) in~\cite{budkov2021theory}) and $\rho_{s}(\psi_0)=q_{+}c_{+s}(\psi_0)+q_{-}c_{-s}(\psi_0)$ is local charge density on the electrode. Note that for the nonpolarizable and nonpolar ions eq. (\ref{cap3}) transforms into eq. (\ref{cap2}). For the case of RTIL with polar or polarizable cation and anion with zero polarizability and dipole moment, the authors investigated the DC profiles for the case of ideal symmetric lattice gas reference model using eq. (\ref{cap3}). They also demonstrated the effect of static polarizability/dipole moment of the cation on the formation of an electric double layer, namely additional attraction of the polarizable cation to the electrode shifts the DC maximum to the negative voltages and increases its magnitude (see Fig. \ref{fig3}). In addition, the competition between the dielectrophoretic and Coulomb forces leads to nonmonotonic behavior of the local concentration of cations on electrode at positive voltages. The latter effect appears only at sufficiently big dipole moments. It turns out that the high polarizabilities prevent the cations from being completely expelled from the positively charged electrode surface. The important result is the crossover from the camel-shaped to the bell-shaped profile taking place at higher ion volume fractions than those predicted by Kornyshev's theory with implicit ion polarizability. The ionic polarizability and dipole moment do not change the asymptotic DC behavior, leading to the dependence $C\simeq (\pm q\varepsilon_{\mp}\varepsilon_0/2v\psi_{0})^{0.5}$, where $\varepsilon_{\pm}=\varepsilon +\alpha_{\mp}/\varepsilon_0v$ are the local dielectric permittivities of the RTIL in the saturated near-surface layer (the signs $\pm$ correspond to the electrode charge signs). The dipole moments $p_{\pm}$ are dropped out of the final expression due to the dielectric saturation effect. It is noted that the obtained asymptotic behavior disagrees with the earlier established result \cite{lauw2009room}, $C\sim |\psi_0|^{-0.6}$, where the RTIL was modeled as a set of dendrimeric ionic particles within the framework of the polymeric lattice self-consistent field theory taking into account the ionic polarizability through the dielectric mismatch of different dendrimer moieties.

{\bf Short-range ionic correlations.} In spite of the achieved progress in adequate description of the DC profiles by taking into account the difference in ionic diameters, short-range interactions, and ionic dielectric properties, the theory is still unable to describe the oscillation behavior of the local concentrations of species and electrostatic potential near the electrode surface at sufficiently low applied voltages (so-called overscreening)~\cite{fedorov2008ionic}, which are determined by the short-range ionic correlations (nonlocal effects).

\begin{figure}[ht]
\subfloat[\centering]{\includegraphics[width=.47\textwidth]{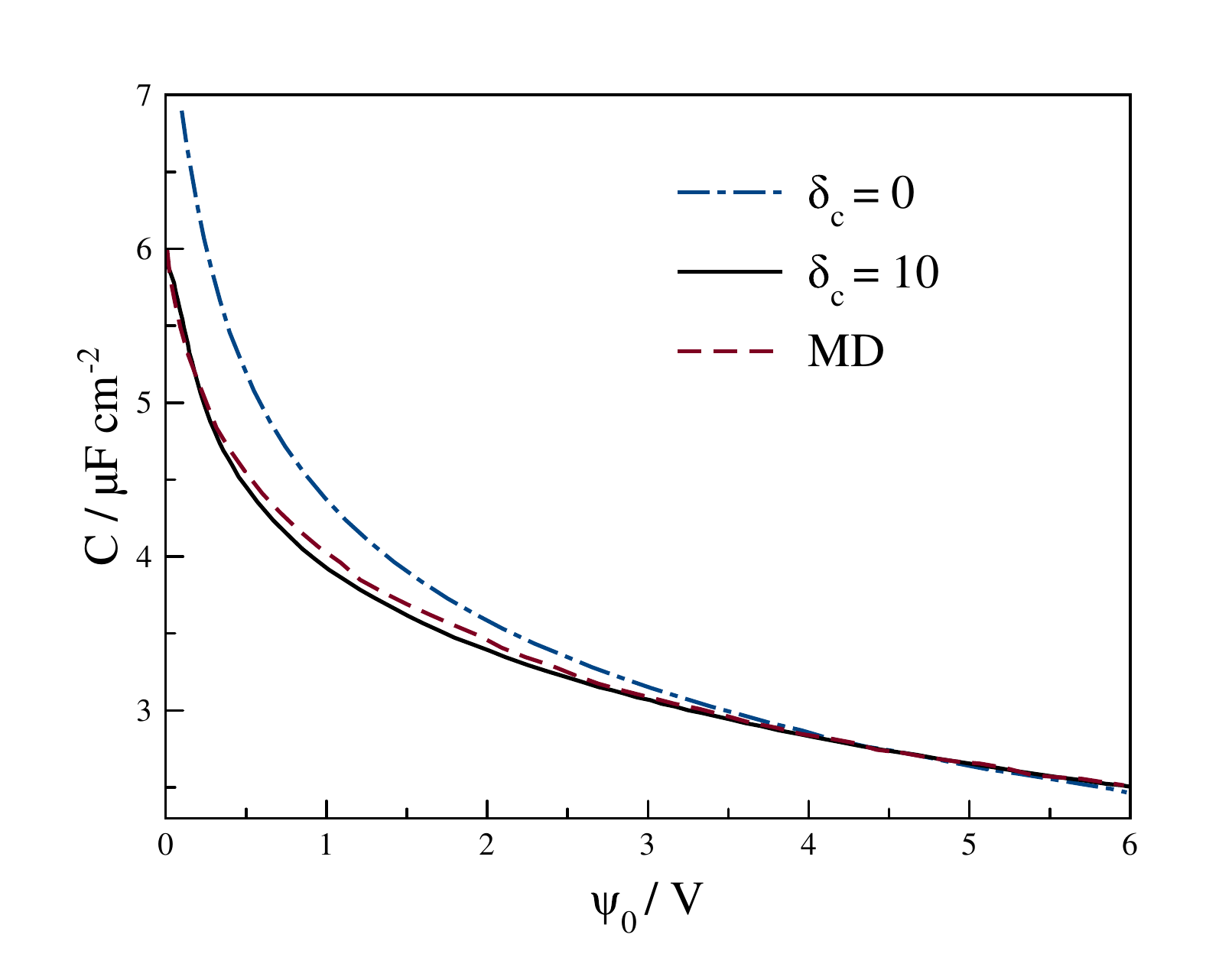}}
\subfloat[\centering]{\includegraphics[width=.47\textwidth]{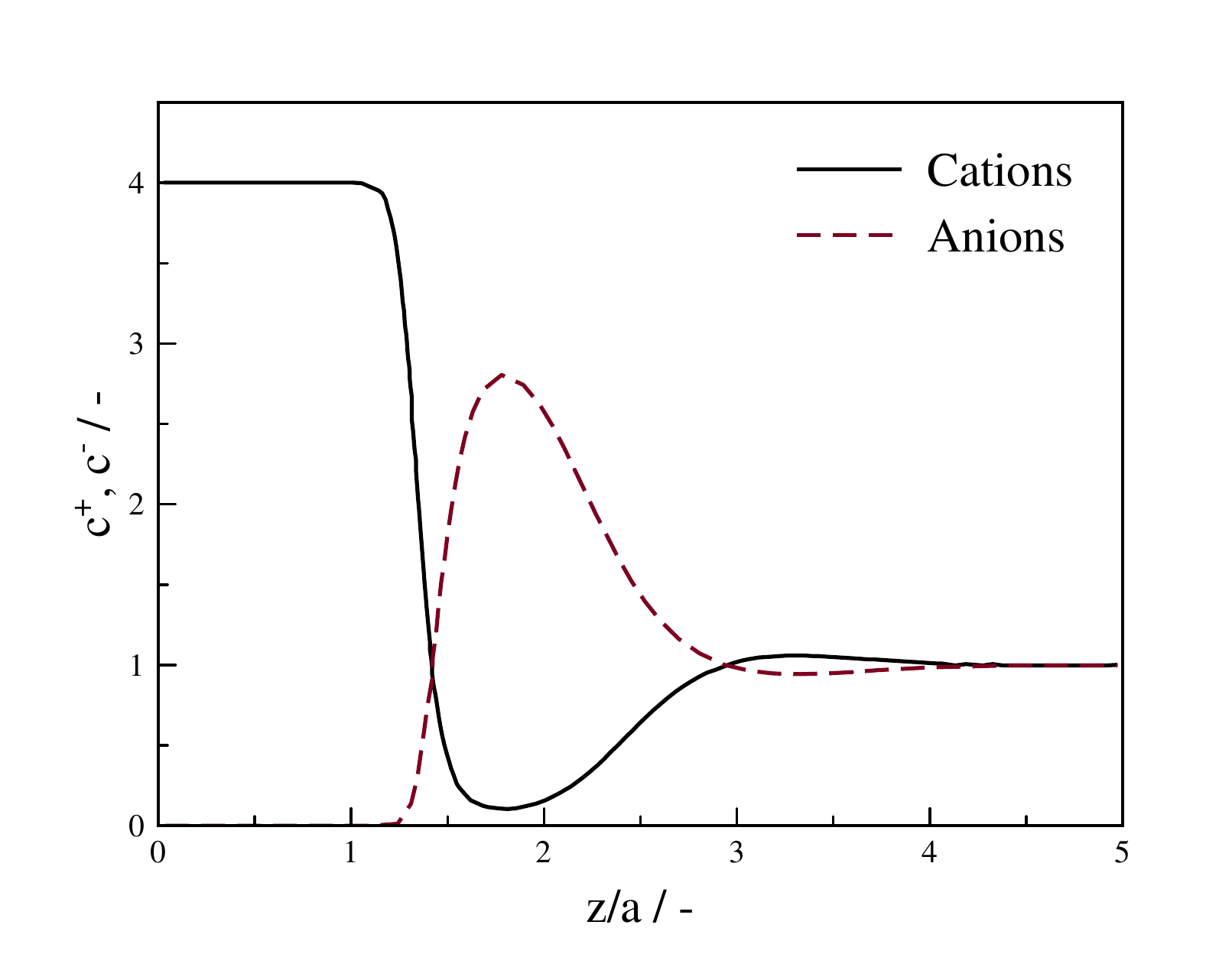}}
\caption{(a) Comparison of differential capacitance profiles calculated within the BSK theory at $\delta_{c}=0$, $\delta_{c}=10$ and obtained in MD simulations \cite{fedorov2008ionic}. (b) Typical profiles of local ionic concentrations calculated within the BSK theory, a is the ion diameter. The ion concentrations were calculated at high voltage $q \psi_0 / k_B T$ = 100. The data is taken from Ref. \cite{bazant2011double}.}
\label{fig4}
\end{figure}

The first to address the problem of taking into account short-range ionic correlations were the authors of paper~\cite{bazant2011double}, who modified Kornyshev's theory by introducing the "nonlocal" contribution $-\varepsilon l_{c}^2{\psi^{\prime\prime}}^2/2$ to the electrostatic free energy density, where $l_c$ is a phenomenological parameter having the meaning of the correlation length -- BSK (Bazant, Storey, Kornyshev) theory. The authors admit that the introduction of the nonlocal contribution is analogous to the implementation of quadratic terms in density gradients in the Cahn-Hilliard theory~\cite{cahn1965phase}. Minimization of the RTIL free energy with an account of the nonlocal term led to the following modified PF equation
\begin{equation}
\label{BSK_eq}
\psi^{\prime\prime}(z)-l_{c}^2\psi^{(IV)}(z)=\frac{2c q}{\varepsilon\varepsilon_0}\frac{\sinh{\beta q\psi(z)}}{1+2cv\left(\cosh\beta q\psi(z)-1\right)}.
\end{equation}
Solving eq. (\ref{BSK_eq}) with the additional boundary condition $\psi^{\prime\prime\prime}(0)=0$, the authors analyzed the DC behavior at different volume fractions, $\gamma$, and "correlation" parameter $\delta_c=l_c/r_{D}$ ($r_D=\sqrt{4\pi\varepsilon \varepsilon_0 k_BTv}/q$ is the Debye radius corresponding to the densest packing of the lattice gas). It was shown that taking the nonlocality into account allowed them to describe the overscreening regime at sufficiently low potentials in qualitative agreement with the molecular dynamics (MD) simulations~\cite{fedorov2014ionic} and nonlocal classical density functional theory (cDFT) results~\cite{wu2011classical}. It was demonstrated that taking into account the nonlocal effects led to qualitative changes in the DC behavior at zero potential (linear capacitance, $C_{lin}$) in comparison with Kornyshev's theory, which predicted the same value as the GC one. Namely, in the BSK theory $C_{lin}\sim \sqrt{2\delta_c+1}/(\delta_c+1)$, i.e. monotonically decreases with the correlation parameter increase. Another important result of the BKS theory is the prediction of the overscreening regime transition at low voltages to the crowding regime at high voltages, where $C\sim |\psi_0|^{-0.5}$, which is also in agreement with the MD simulations. Although the phenomenological BSK theory can successfully describe the behavior of the DC and potential profiles, it is unable to expose the nature of the correlation length, $l_c$. As it has been lately shown, the occurrence of the nonlocal terms in the electrostatic free energy can be explained by the presence of quadrupolar ionic clusters in the RTIL bulk~\cite{avni2020charge}. In this case, the correlation length has the meaning of quadrupolar length~\cite{slavchov2014quadrupole,slavchov2014quadrupole_2,budkov2019statistical,budkov2020statistical}. Note that the dipolar clusters (ionic pairs), as it was shown in papers~\cite{zhang2020enforced,avni2020charge}, make a contribution to the RTIL dielectric constant. Note that the presence of ionic clusters in an RTIL is proved by the MD simulations data~\cite{feng2019free}. Another conceptual problem of the BKS theory is the postulation of the boundary condition $\psi^{\prime\prime\prime}(0)=0$, which cannot be strictly justified from first principles. An interesting question is why the BSK model and the MD simulations, shown in Fig. \ref{fig4}, agree with each other, despite the incorrect boundary condition. The problem of the boundary conditions in the mean-field-like theories stems from the fact that the microscopic EDL theory must be essentially nonlocal, while the account of the gradient contributions to the electrostatic free energy density is associated with the truncation of the power series in the wave vector of the dielectric function, $\varepsilon(\bold{k})$, at the quadratic term~\cite{avni2020charge,bazant2011double,budkov2020statistical}. We note that the authors of papers~\cite{de2020continuum,misra2019theory} made an attempt to overcome the ambiguity of the boundary condition choice for electrolyte solutions on electrified surface and in confinement.

Another interesting theory was formulated in work~\cite{blossey2017structural}. Utilizing the formalism based on the Legendre transformation for the case of an arbitrary RTIL reference system, the authors introduced short-range ionic correlations at the level of the Cahn-Hilliard approximation. The authors obtained the oscillating behavior of the electrostatic potential and showed that the equation for the electrostatic potential in the linear approximation has a structure close to that of the equation in the BSK theory. Similar results were obtained within the phenomenological theory~\cite{ciach2018simple}. It is worth noting the theory \cite{downing2018differential}, where the authors took into account the short-range ionic correlations at the level of the Cahn-Hilliard approximation and specific interactions of the ions at the level of Bragg-Williams approximation, and studied the IL stability in the EDL. A similar theory was proposed in~\cite{cruz2018electrical}, where the authors studied the DC behavior for the RTIL-solvent mixture in the region of demixing transition. Paper \cite{huang2018confinement} discusses the electrostatic screening in RTILs under nanoconfinement. De Sousa et al.~\cite{de2020interfacial} proposed a simplified version of the nonlocal cDFT, based on the hard spheres equation of state, which allowed the authors to describe the overscreening and crowding regimes, as well as the crossover between them. In contrast to the previous analytical theories, this one predicted the formation of discrete charged layers on the electrode surface and slowly damped oscillations of the local ionic concentrations with a wave length of the order of the ionic diameter. An important point of the theory is quantitative agreement with the MD simulation. It is interesting to note that the equation for the electrostatic potential in the linear approximation also has a structure similar to that of the equation from the BSK theory. We would like to note that refs. \cite{jiang2011density,wu2011classical,forsman2011classical} take into account the short-range ionic correlations within cDFT. Overcharging is explicitly demonstrated in the work \cite{wu2011classical} and in ref. \cite{forsman2011classical}, it is at least implicitly demonstrated from the oscillating net charge density profiles.

{\bf Concluding remarks and prospects.} In this paper, we discussed the most important theoretical results obtained since the publication of Kornyshev's paper~\cite{kornyshev2007double} in the field of EDL description at the RTIL-metal interface. Despite the significant progress, some fundamental problems remain unsolved. First, as is well known~\cite{naji2013perspective}, the mean-field theory describes the behavior of an RTIL and an electrolyte solution in the region of rather weak electrostatic interactions, completely neglecting the electrostatic correlations of the ions~\cite{levin2002electrostatic}. Significant advances have been lately made for the electrolyte solutions with polyvalent ions~\cite{misra2019theory,buyukdagli2020schwinger}. To the best of our knowledge, nowadays there is no such theory for RTILs. The second important problem is explicit account of the electrons in the metal electrode. This will enable us to consider the chemical specificity of the electrode. A successful attempt has been recently made to take into account the quantum specificity of the electrode in the framework of the combination of the quantum and classical density functional theories for electrolyte solutions~\cite{huang2021hybrid}. However, such theory for the RTILs on charged electrodes is still nonexistent. Finally, all the discussed above theoretical models do not take into account an electrode roughness, as would be present on many real surfaces, e.g. platinum. It is important to understand how the electrode roughness influences the differential capacitance. We hope that solving these problems is a matter of the nearest future.

{\bf Acknowledgements.} YAB thanks the Russian Science Foundation (Grant No. 21-11-00031) for financial support. The authors thank Nikolai Kalikin for assistance in manuscript preparation.


\bibliographystyle{elsarticle-num.bst}
\bibliography{lit}

\end{document}